\documentclass{elsart}

\usepackage{natbib}

\usepackage{graphicx}

\usepackage{amssymb}

\def\apj #1 #2 #3 {#1, ApJ, {\bf #2}, #3}
\def\apjl #1 #2 #3 {#1, ApJ, {\bf #2}, L#3}
\def\apjs #1 #2 #3 {#1, ApJS, {\bf #2}, #3}
\def\aap  #1 #2 #3 {#1, A\&A, {\bf #2}, #3}
\def\mnras #1 #2 #3 {#1, MNRAS, {\bf #2}, #3}
\def\pra #1 #2 #3 {#1, Phys.~Rev.~A., {\bf #2}, #3}
\def\prb #1 #2 #3 {#1, Phys.~Rev.~B., {\bf #2}, #3}
\def\prc #1 #2 #3 {#1, Phys.~Rev.~C., {\bf #2}, #3}
\def\prd #1 #2 #3 {#1, Phys.~Rev.~D., {\bf #2}, #3}
\def\pre #1 #2 #3 {#1, Phys.~Rev.~E., {\bf #2}, #3}
\def\prl #1 #2 #3 {#1, Phys.~Rev.~Lett., {\bf #2}, #3}
\def\plb #1 #2 #3 {#1, Phys.~Lett.~B., {\bf #2}, #3}
\def\science #1 #2 #3 {#1, Science., {\bf #2}, #3}
\def\nature #1 #2 #3 {#1, Nature., {\bf #2}, #3}
\def\nphysa #1 #2 #3 {#1, Nucl.~Phys.~A., {\bf #2}, #3}
\def\nphysb #1 #2 #3 {#1, Nucl.~Phys.~B., {\bf #2}, #3}
\def\nphysbs #1 #2 #3 {#1, Nucl.~Phys.~B.~Suppl., {\bf #2}, #3}

\def\h#1{\hbox{${}^{#1}$H}}

\def\h502{\hbox{$ h^{2}_{50}$}}

\def\la{\mathrel{\mathpalette\fun <}}
\def\ga{\mathrel{\mathpalette\fun >}}
\def\fun#1#2{\lower3.6pt\vbox{\baselineskip0pt\lineskip.9pt
  \ialign{$\mathsurround=0pt#1\hfil##\hfil$\crcr#2\crcr\sim\crcr}}}

\begin{document}

\begin{frontmatter}


 \title{New Classes of Cosmic Energy\\ and Primordial Black-Hole Formation}
 \author{K. Ichiki\thanksref{label1}\thanksref{label2}\corauthref{cor1}}
 \ead{ichiki@th.nao.ac.jp}
 \author{M. Orito\thanksref{label1}\corauthref{cor2}}
 \ead{orito@nr.titech.ac.jp}
 \author{T. Kajino\thanksref{label1}\thanksref{label2}\corauthref{cor1}}
 \ead{kajino@nao.ac.jp}


 \address[label1]{National Astronomical Observatory, 2-21-1, Osawa, Mitaka,
 Tokyo 181-8588,Japan}
 \address[label2]{University of Tokyo, Department of Astronomy, 7-3-1
 Hongo, Bunkyo-ku, Tokyo 113-0033, Japan}
 \corauth[cor1]{Institute for Nuclear Theory, University of Washington,
 Seattle, WA 98195-1550, U.S.A}
 \corauth[cor2]{Present address: Research Laboratory for Nuclear
 Reactors, Tokyo Institute of Technology 2-12-1 Ookayama, Meguro-ku,
 Tokyo, 152-8550, JAPAN}


\begin{abstract}
 It has recently been suggested that the formation of horizon-size
 primordial black hole (PBH) from pre-existing density fluctuations is
 effective during the cosmic QCD phase transition.  
 In this Letter we discuss the dependence of PBH formation on effective
 relativistic degrees of freedom, $g_{\rm eff}$ during the cosmic QCD phase
 transition. 
 Our finding is important in the light of recent
 cosmological arguments of several new classes of cosmic energy that
 appear from universal  neutrino degeneracy, quintessential inflation,
 and dark radiation in  brane world cosmology. Extra-energy component
 from the standard value  in these new cosmological theories is
 represented as an effective  radiation in terms of $g_{\rm eff}$. We
 conclude that the PBH formation during QCD  phase transition becomes
 more efficient if negative extra-component of  the cosmic energy is
 allowed  because of the increase of the duration of the QCD phase
 transition, which leads to smaller mass scale of PBHs. This suggests
 larger probability of finding more  PBHs if the dark radiation exists
 as allowed in the brane world cosmology.  
\end{abstract}

\begin{keyword}
Cosmology \sep Quark-gluon plasma \sep Black holes
\PACS 98.80 \sep 12.38.M \sep 97.60.L
\end{keyword}

\end{frontmatter}





\section{Introduction}
First argument of the PBH formation was presented by Carr and Hawking
\cite{carr.hawking}, in which 
the key concept of the PBH formation is that the three different length
scales of the particle horizon, the Schwarzschild radius, and the Jeans
length of any overdense region at radiation-dominated epoch are of the
same order of magnitude. 
Once an initially superhorizon-size density fluctuation crosses the
particle horizon, competition between self gravity and centrifugal
pressure force determines whether such a fluctuation collapses into black
hole or not\cite{hawking,carr}.
If the self gravity overcomes the pressure force, then the fluctuation will
form a black hole with the order of horizon mass.
Jedamzik \cite{jedamzik} found that, during cosmic QCD phase transition,
there is almost no
pressure response on adiabatic compression of fluctuation in the mixed
phase of quark-gluon plasma and hadron gas.
This could occur because the enhanced energy density due to the
compression of the fluctuations will turn into the high
energy quark-gluon phase, instead of increasing centrifugal pressure
force. 
As a result the overdense region from initially existing fluctuations
tends to collapse efficiently to form the PBH during the QCD 
epoch when horizon mass scale is around 1$M_\odot$.
PBH mass spectrum is most likely dominated by
the horizon mass at the phase transition epoch \cite{niemeyer}.
Green and Liddle \cite{green2} verified that this speculation turns out
to be a good approximation in their detailed studies of the PBH mass spectrum.

These theoretical speculations have been studied more quantitatively in
1D general-relativistic hydrodynamic calculations \cite{niemeyer2},
where the required critical overdensity parameter, $\delta_{hc}\equiv
\delta \rho/\rho$ , was found to be $\delta_{hc} \sim 0.7$ for the PBH
formation.
Similar sophisticated calculations \cite{jedamzik2} have also shown 
numerically that $\delta_{hc}$ is even as low as $\delta_{hc} \sim 0.54$ by 
taking account of the dynamics of the QCD phase transition in
supercooling phase by the use of EOS 
in the MIT bag model for the mixed phase.

Recently an interesting cosmological proposal has been made that our
universe is occupied fractionally by a new class of extra-energy 
component in addition to the ordinary matter, known as
``quintessence''. The quintessential scalar field causes the
accelerating universal expansion at present \cite{perlmutter,garnavich}
and it also has
a possibility of making significant contribution to the total energy
density at radiation dominated epoch \cite{ratra,steinhardt,bean,yahiro}.
Such an extra-energy density is interpreted as to increase 
relativistic degrees of freedom, $g_{\rm eff}$,  and thereby causes more
rapid expansion rate at the same temperature.
Similar effect is expected in the neutrino-degenerate universe models
\cite{orito}, and the universal lepton asymmetry was critically studied 
in the context of constraining the primordial nucleosynthesis and cosmic
microwave background anisotropies.
On the other hand, another attempt has been made quite recently to
understand the Einstein gravity in the light of the unified theory
\cite{shiromizu,mukoyama,binetruy,ida}. 
Five dimensional brane world cosmology leads to an effective theory of
gravity, which is motivated by the superstring theory or M-theory, which
is a strong candidate for the unified theory.
In this scenario the ``standard'' Friedmann equation should be slightly
modified by adding several new terms which arise from the extra dimensions.
Several authors \cite{mukoyama,binetruy,ida} showed that there is a new
class of effective energy among them which diminishes in proportion to 
$a^{-4}$ called ``dark radiation'', where $a$ is the cosmic scale factor.
The interesting point is that this dark radiation term could be even
negative because it originates purely from fifth dimensional
geometry. If this is the case, negative dark radiation causes the cosmic
expansion slower. We can therefore interpret this effect as a decrease
in effective degrees of freedom, $g_{\rm eff}$.

The purpose of the present Letter is to study how strongly the change of the
relativistic degrees of freedom $g_{\rm eff}$ affects the PBH formation during QCD phase transition.
The energy density of the universe gets larger for larger $g_{\rm eff}$ at
the same temperature, but the duration of QCD phase transition becomes
shorter. These two effects operate in opposite direction to form PBHs.
Therefore, it is not a trivial subject since many physical quantities may be
affected in a complicated manner by this parameter $g_{\rm eff}$.
In order to clarify its effects we use a simple model for the 
numerical analysis instead of performing complicated hydrodynamic calculations.
We then try to find the parameter dependence of the typical mass scale and the 
amount of PBHs which are formed during the QCD phase transition.

\section{QCD Phase Transition}
We first briefly  review important quantities involved in the PBH
formation during the QCD phase transition.
Horizon mass at the QCD epoch is written as
\begin{equation}
M_h(T)=0.8M_{\odot}\left(\frac{T}{100\rm{MeV}}\right)^{-2}
\left(\frac{g_{\ast}}{51.25}\right)^{-\frac{1}{2}},
\end{equation}
where $g_{\ast}$ is the number of relativistic degrees of freedom, and T
is the cosmic temperature.
This equation shows that the horizon mass at the QCD phase transition is
compatible to typical mass scale of MACHOs \cite{alcock}.

Recent progress in lattice QCD calculations and accelerator experiments
have provided with rich information of the QCD phase transition.
We employ the MIT bag model to express the
quark-gluon plasma phase \cite{mathews}. In this model the energy densities of
quark-gluon plasma ($\rho_{\rm qg}$) and hadron gas ($\rho_{\rm h}$) and
the pressures ($p_{\rm qg}$ and $p_{\rm h}$) respectively are given by
\begin{equation}
\rho_{\rm qg}=\frac{\pi^2}{30}g_{\rm qg}T^4+B,
\hspace{1cm}
\rho_h(T)=\frac{\pi^2}{30}g_{\rm h}T^4,
\end{equation}
\begin{equation}
p_{\rm qg}=\frac{1}{3}\rho_{\rm qg}-B,
\hspace{1.7cm}
p_h(T)=\frac{1}{3}\rho_h,
\end{equation}
where $g_{\rm qg}$ and $g_{\rm h}$ are the numbers of degrees of freedom of
the quark-gluon plasma and hadron gas phases,
and B is the Bag constant.
The condition of pressure balance between the two phases, 
$p_h(T_{QCD})=p_{\rm qg}(T_{QCD})$, for a first order phase transition yields
\begin{equation}
B=\frac{\pi^2}{90}(g_{\rm qg}-g_{\rm h})T_{\rm QCD}^4,
\end{equation}
\begin{equation}
L=\rho_{\rm qg}(T_{\rm QCD})-\rho_h(T_{\rm QCD})=4B,
\end{equation}
where $L$ is the latent heat of the phase transition.
This implies that the intensity of the first order QCD phase
transition depends on the difference between the two degrees of
freedom before ($g_{\rm qg}$) and after ($g_{\rm h}$) the phase
transition.

Duration of the phase transition is expressed by the change of the
cosmic scale factor during the phase transition.
From the conservation law of entropy we estimate the rate
\begin{equation}
\frac{a_2}{a_1}=\left(\frac{g_{\rm qg}}{g_{\rm h}}\right)^{\frac{1}{3}}\approx 1.44,
\end{equation}
where $a_1$ and $a_2$ are the scale factors at the beginning and the end of 
the QCD phase transition, respectively.
We can learn from this that the duration of the phase transition is
comparable to the Hubble time at the QCD epoch, and it is enough for
fluctuations to collapse gravitationally.

\section{Analysis}
\subsection{Theoretical Model}

For the numerical analysis of the PBH formation we mainly follow the
method developed by Cardall and Fuller \cite{cardall}.
The condition for the gravitational collapse of PBH is that
the self gravity is stronger than the centrifugal pressure force, which is
expressed as
\begin{equation}
G\rho^2S^5_{\rm coll} \ga pS^3_{\rm coll},
\end{equation}
where $\rho$ is energy density, p is pressure, and $S_{\rm coll}$ is the
size of an overdense region at the   turn around. 
This condition becomes
\begin{equation}
S_{\rm coll} \ga \sqrt{\frac{w}{G\rho}} \approx R_J 
\end{equation}
where $w=\frac{p}{\rho}=\frac{1}{3}$ in the radiation dominated era and
$R_J$ is the Jeans length. Equation (8) is a well known condition for the
gravitational collapse.
This condition is also expressed, using the density contrast
$\delta_{\rm hc}$ at horizon crossing \cite{carr}, as
\begin{equation}
\delta_{\rm hc} \ga w.
\end{equation}
This means that  fluctuations having density
contrast above $\frac{1}{3}$ at the horizon crossing can collapse to
form PBH at radiation dominated era.

We need modification of Eq.(9), taking account of the effects of the QCD
phase transition.
In order to proceed quantitative discussions, we introduce the quantity
$f$ and modify the above condition by
\begin{equation}
\delta_{hc} \ga w(1-f),
\end{equation}
\begin{equation}
f=\frac{S^3_2-S^3_1}{S^3_{\rm coll}},
\end{equation}
where $S_2$ and $S_1$ are the sizes of initially expanding fluctuation at
the beginning and the end of the phase transition, and $S_{coll}$ is the
size at turn around \cite{cardall}.
This quantity $f$ works for softening the equation of
state $w$ and takes the value,
$0\la f \leq 1-\left(\frac{S_1}{S_{coll}}\right)^3$.
When $f$ is the minimum value, $f=0$, the fluctuation is not affected
by the QCD phase transition at all, and Eq.(10) turns out to be the
simplest case of Eq.(9).
When $f$ is the maximum value, $f=1-\frac{S_1}{S_{\rm coll}}$, the
fluctuation has been affected the most by the phase transition because
the fluctuation starts collapsing early enough so that the overdense
region includes the mixed phase of quark-gluon plasma and hadron gas.
In the latter case the system does not feel any effective
pressure response.
$S_2$ and $S_1$ are calculated from the energy density of the background
and the energy contrast of the fluctuation  when the fluctuation crosses
the particle horizon. We solve the flat Friedmann equation for the background
\begin{equation}
\left(\frac{da}{dt}\right)^2 = \frac{8 \pi G}{3} \bar \rho a^2,
\end{equation}
and the closed Friedmann equation for the fluctuation
\begin{eqnarray}
\left(\frac{dS}{d\tau}\right)^2 &=& \frac{8 \pi G}{3} \rho S^2 -k,\\
&=& \frac{8 \pi G}{3}\bar \rho_h R_h^{3(1+w)}
\left(\frac{1+\delta}{S^{1+3w}}-\frac{\delta}{R_h^{1+3w}}\right).
\end{eqnarray}
In these equations, we assume that the fluctuation has the top hat
profile $\rho_h = \bar\rho_h(1+\delta)$, where $\delta$ means the density
contrast at horizon crossing.
We normalize the density evolution and fix the gauge by setting
$\tau_h=t_h$, $S_h=R_h$, and $(dS/d\tau)_h=(dR/dt)_h$ when the
fluctuation crosses the horizon.

We show the quantity $w(1-f)$ as the function of $\delta$ in
Fig.1, from which we can show the effect of QCD phase transition on
the PBH formation. 
Here, $x \equiv \frac{\bar \rho_h}{\rho_1}$, ${\bar\rho_h}$ is the mean energy
density when the fluctuation crosses the horizon, and $\rho_1$
is the mean energy density above which the universe is in pure quark-gluon
plasma phase. This variable $x$ indicates when the fluctuation under
consideration crosses the particle horizon.

\begin{figure}
\centering
\includegraphics[width=1.0\textwidth]{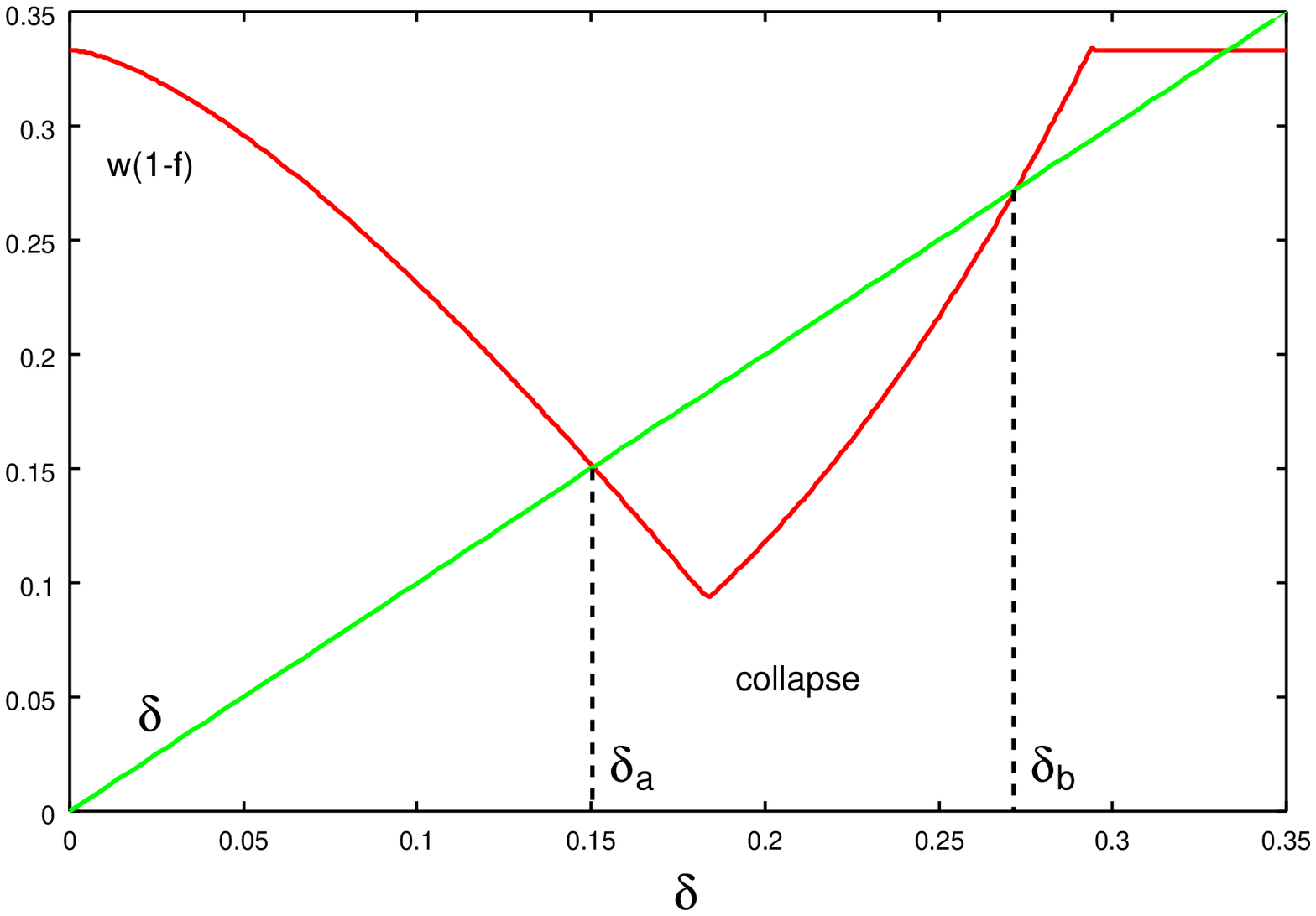}
\caption{\label{fig:fig1}
 Modification of collapsing condition for $x={\bar \rho_h}/\rho_1=15$. $x=15$ corresponds to a
 fluctuation with mass scale $M_{\rm PBH} \sim 0.2M_{\odot}$ \cite{cardall}.
 $w(1-f)$, for $w=\frac{1}{3}$, and $\delta$ are shown as indicated.
 Collapse region $\delta_a \le \delta \le \delta_b $satisfies the
 collapsing condition of Eq.(10).}
\end{figure}
In order to connect the physics of the QCD phase transition with the
relativistic number of degrees of freedom, we introduce
effective background degrees of freedom,$g_{\rm eff}=g_{\rm bg}+\Delta
g$, where $g_{\rm bg}\approx 14.25$ ($\gamma$, $e^{\pm}$,
$\mu^{\pm}$, three kinds of neutrinos in the standard case), and $\Delta
g$ is for the extra-component at the QCD phase
transition epoch as to be discussed in the next section. 
The numbers of degrees of freedom of the hadron phase and
quark-gluon phase are respectively given by, $g_{\rm h}=g_{\rm eff}+3$
(pions), and $g_{\rm qg}=g_{\rm eff}+ 37$ (21 for quarks and 16 for gluons).
We thus obtain the relation between the parameter
$x$ and the number of effective degrees of freedom $g_{\rm eff}$ for horizon mass scale $M$,
\begin{equation}
x=\frac{\bar \rho_{\rm h}}{\rho_1}=\frac{3g_{\rm eff}+145}{153.75}
\left(\frac{M}{0.87M_{\odot}}\right)^{-2}.
\end{equation}

\subsection{Present Day PBH Density}
We consider the significance of the PBHs formed in the QCD phase
transition in this sub-section.
Present cosmological density parameter of PBHs formed in the QCD phase
transition is represented by
\begin{equation}
\Omega_{\rm PBH}h^2 = 4.8\times 10^6 g(T)\varepsilon(T)
\left(\frac{T_0}{2.73 \rm{K}}\right)^3 
\left(\frac{T}{100 \rm{MeV}}\right),
\end{equation}
where $h$ is the Hubble parameter in units of 100 km/s/Mpc, $g(T)$ is the 
relativistic degrees of freedom, and 
$\varepsilon(T)$ is the fraction of the radiation energy density
which is converted into black holes at temperature T. $\varepsilon(T)$ is
defined by
\begin{equation}
\varepsilon(T)=\int_{\delta_a}^{\delta_b}F(\delta,T)d\delta,
\label{epsilon}
\end{equation}
where $F(\delta,T)$ is the probability function of a horizon volume
fluctuation to have overdensity parameter $\delta$ \cite{green}
\begin{equation}
F(\delta,T)=\frac{1}{\sqrt{2\pi} \sigma(M)}\exp\left(-\frac{\delta^2}{2\sigma^2(M)}
\right),
\end{equation}
\begin{equation}
\sigma(M)=9.5\times 10^{-5}\left(\frac{M}{10^{22}M_{\odot}}\right)^{(1-n)/4}.
\label{sigma}
\end{equation}
In Eq. (\ref{sigma}) $n$ is the spectral index of primordial density 
fluctuations.
We consider the fluctuation which has overdensity larger than $\delta_a$
[see Eq. (\ref{epsilon})]
and can collapse into black holes, so that it satisfies  Eq.(10) for a
given $x$ (See Fig.1).
It is to be noted that even a small fraction $\varepsilon$ can result in
the significant amount of cosmological energy density at present.
From Eq.(15) we can estimate the mass of the PBH and the variance
$\sigma$ for a given $x$ and n. We can also
calculate the efficiency $\varepsilon$ and $\Omega_{\rm PBH}$ using
Fig.1 and Eqs.(15)-(19).

\section{Results and Discussions}
If the universe is modeled to have a new class of extra-energy component
, it leads to increase relativistic degrees of
freedom, $g(T)$.
In the neutrino-degenerate universe models \cite{orito}, additional degrees
of freedom are written as
\begin{equation}
\Delta g = \sum_\alpha \left(\frac{T_{\gamma_{\alpha}}}{T_\gamma}\right)^4\left[\frac{30}{7}\left(\frac{\xi_\alpha}{\pi}\right)^2
+\frac{15}{7}\left(\frac{\xi_\alpha}{\pi}\right)^4\right],
\end{equation}
where $\alpha$'s correspond to $e$-,$\mu$- and $\tau$-neutrino species,
and $\xi$ is the neutrino chemical potential divided by the neutrino
temperature, $\xi_\alpha \equiv \mu_\alpha/T_\alpha$.
We note that slightly non-zero chemical potential is
favored in recent BBN analysis with better
goodness of fit to the recent data of CMB anisotropies \cite{orito},
which constrains $\Delta g \la 4.72$.

Another class of extra-energy component appears in quintessential
inflationary model \cite{ratra,steinhardt,bean,yahiro}
\begin{equation}
\Delta g = g_{\rm Q} = g_{\rm B}\frac{\rho_{\rm Q}}{\rho_{\rm B}},
\end{equation}
where $\rho_B$ is the background photon energy density, 
$\rho_B=\rho_{\gamma0}(1+z)^4(g_0/g(z))^{1/3}$, and $\rho_{\rm Q}$ is the
energy density of quintessence field,$\rho_{\rm Q}=\dot{Q}_{\rm
QCD}^2/2+V(Q_{\rm QCD})$.
$Q_{\rm QCD}$ and $\dot{Q}_{\rm QCD}$ are the quintessence scalar field value
and its time derivative at the QCD phase transition.
$\Delta g$ is constrained to be $\la 0.2$ so as to satisfy the BBN and
CMB analyses \cite{yahiro}.

In a recent theory of brane world cosmology
\cite{shiromizu,mukoyama,binetruy,ida}, completely a new class of cosmic
energy, called dark radiation, emerges from extended Friedman equation.
Its extra-component is expressed by
\begin{equation}
\Delta g = g_{\rm DR} = g_{\rm B} \frac{\rho_{\rm DR}}{\rho_{\rm B}}
\mbox{,\hspace{0.3cm}}
\label{dark}
\end{equation}
where $\rho_{\rm DR}$ is the dark-radiation term, $\rho_{\rm
DR}=\frac{3\mu}{8\pi G}(1+z)^4$, and $\mu$ is a constant which comes
from the electric Coulomb part of the five-dimensional Weyl tensor
\cite{shiromizu}.
We note that the dark radiation
can be negative depending on the sign of $\mu$.
This is in remarkable contrast to Eqs.(20) and (21) which
are always positive.
The BBN and CMB constraints \cite{ichiki} on $\Delta g$ in the brane
world cosmology is
$-0.47 \la \Delta g \la 0.12$ (at the $2\sigma$ C.L.).
Note that only the BBN constraint allows $-4.65\la \Delta g$.

\begin{figure}
\centering
\rotatebox{-90}{\includegraphics[width=0.65\textwidth]{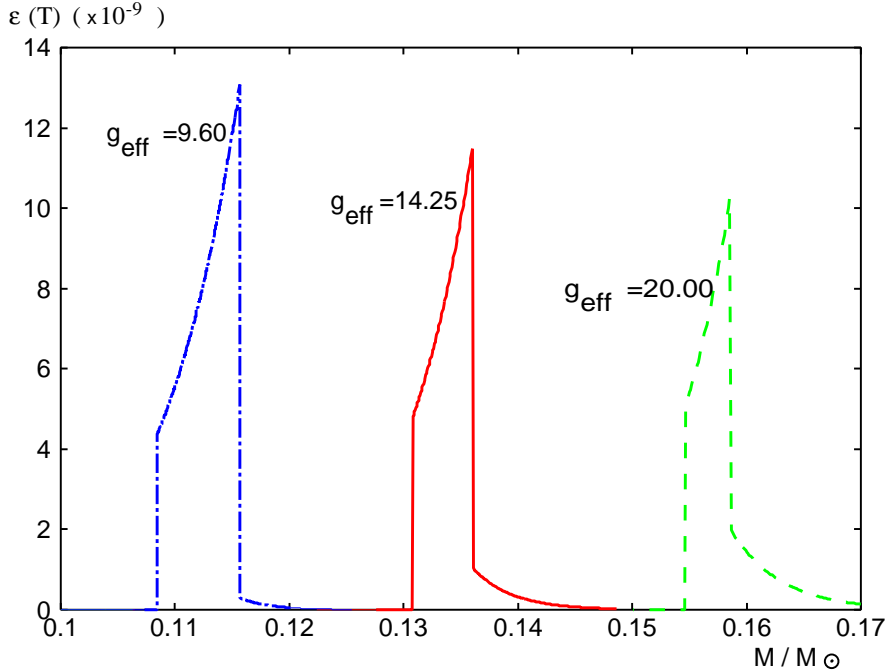}}
\caption{\label{fig:epsilon}
The fraction of the radiation energy density converted into black holes
 in three cases of $g_{\rm eff}$ values as a function of the mass of PBH.
This function is steeply peaked for any $g_{\rm eff}$, which suggests that PBH forms at a
 particular epoch for almost single horizon mass corresponding to a
 peaked $M$. In this figure we
 have tuned spectral index so that $\Omega_{\rm PBH}\approx 1$ at
 present for the sake of illustration.}
\end{figure}

We show the calculated results for $g_{\rm eff}=9.60, 14.25,$ and
$20.00$ in the Figs.2 and 3.
$g_{\rm eff}=9.60$ corresponds to $\rho_{\rm extra}/\rho_{\rm
B} = -0.27$ just after the QCD phase transition, which is
the lowest value deduced from only the BBN constraint in brane world cosmology
\cite{ichiki}. $g_{\rm eff}=14.25$ is for the standard case, and
$g_{\rm eff}=20.00$ corresponds to 
$\rho_{\rm extra}/\rho_{\rm B}=0.33$
,which is a likely case in the quintessential inflationary scenario
\cite{yahiro}.
Refer to ref. \cite{yahiro} for more
critical discussions of the observational constraints on $\rho_{\rm
extra}/\rho_{\rm B}$ in quintessence scalar fields.

The key results are summarized as follows.
Fine tuned Gaussian
blue spectra fluctuation is needed with spectral index $n \approx 1.354 \pm
0.005$, $1.377 \pm0.005$, $1.395\pm 0.005$ in the cases of $g_{\rm eff}=9.60$
, $14.25$, and $20.00$ respectively, in order to have significant PBHs energy
density $\Omega_{\rm PBH}$ at present, i.e. $0.01 \leq \Omega_{\rm PBH}
\leq 1$.

Typical mass scales of
PBH are $0.116M_{\odot}$, $0.136M_{\odot}$, and $0.159M_{\odot}$
for $g_{\rm eff}=9.60, 14.25$ and $20.00$, respectively, where three
mass scales refer to maximum  $\varepsilon (T)$'s.
(Eq.(16) and Fig.2). 
The spectral indices as shown in Fig.3 are compatible with those inferred
from the data of COBE observations, 
but larger than the best fit analysis of the combined different cosmological
data sets \cite{wang}. These indices are slightly larger than the values
constrained from the evaporation of PBHs by Hawking
radiation in much smaller mass scale \cite{green}.
Increasing the effective number of degrees of freedom $g_{\rm eff}$
causes the shorter duration of the QCD phase transition because the
universal expansion obeys
\begin{equation}
\frac{a_2}{a_1}=\left(\frac{g_{\rm qg}}{g_{\rm h}}\right)^{1/3}
               =\left(\frac{g_{\rm eff}+37}{g_{\rm eff}+3}\right)^{1/3},
\end{equation}
instead of Eq.(6). Accordingly, the duration of the mixed phase also
becomes shorter, making less efficient PBH formation (Fig.3).
The typical mass scale of the PBHs, however, increases with increasing
$g_{\rm eff}$ (Fig.2). 
This is because the fluctuations that are most effectively influenced
by the QCD phase transition enter the horizon at later time with
increasing number of degrees of freedom.

\begin{figure}
\centering
\rotatebox{-90}{\includegraphics[width=0.6\textwidth]{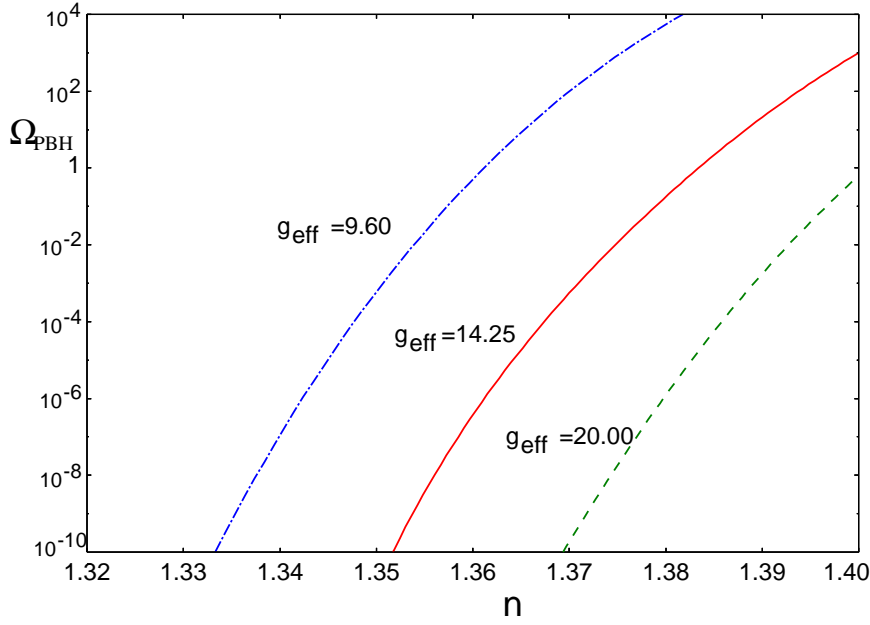}}
\caption{\label{fig:epsart}Density parameter of PBHs $\Omega_{\rm PBH}$at present, as a function of the spectral index $n$ of the primordial density fluctuations.}
\end{figure}

\section{Conclusion}
We studied the formation mechanism of PBHs during the QCD phase
transition, being
motivated by the cosmological interest in MACHOs as a dark matter and
also new classes of cosmic extra-energy components for the degenerate
neutrinos, dark energy, and dark radiation.
In the present
paper we take account of the effects of these extra-components
by means of parameterizing their effect as an increase or
decrease in effective relativistic degrees of freedom.
We conclude that the PBH formation during QCD phase transition becomes
more efficient with an negative extra-component because of the increase
of the duration of the QCD phase transition. We also found that in such
a case the typical mass
scale of PBH becomes smaller. These results suggest larger probability of
finding more PBHs that have typically sub-solar mass $M\sim 0.1M_\odot$
if the dark radiation exists as allowed in the brane
world cosmology.

\end{document}